\documentclass[12pt]{iopart}
\usepackage{bbm}
\usepackage{graphicx}
\usepackage{iopams}
\newcommand{\ket}[1]{\vert #1 \rangle}
\newcommand{\bra}[1]{\langle #1 \vert}

\newcommand{\bmsigma}{\boldsymbol \sigma}
\newcommand{\bmSigma}{\boldsymbol \Sigma}

\newcommand{\bmA}{\boldsymbol A}
\newcommand{\bmB}{\boldsymbol B}
\newcommand{\bmC}{\boldsymbol C}
\newcommand{\bmS}{\boldsymbol S}
\newcommand{\bmLambda}{\boldsymbol \Lambda}

\newcommand{\calA}{{\cal A}}
\newcommand{\calB}{{\cal B}}

\newcommand{\sfx}{{\sf x}}
\newcommand{\sfy}{{\sf y}}
\begin{document}
\title{Squeezed Fock state by inconclusive photon subtraction}
\date{\today}
\author{Stefano Olivares\footnote{Stefano.Olivares@mi.infn.it} 
and Matteo G. A. Paris\footnote{Matteo.Paris@fisica.unimi.it}}
\address{Dipartimento di
Fisica, Universit\`a degli Studi di Milano, Italia}
\begin{abstract}
We analyze in details the properties of the conditional state recently
obtained by J.~Wenger {\em et al.}~[Phys.~Rev.~Lett.~{\bf 92}, 153601
(2004)] by means of inconclusive photon subtraction (IPS) on a squeezed
vacuum state $S(r)\ket{0}$. The IPS process can be characterized by two
parameters: the IPS transmissivity $\tau$ and the photodetection quantum
efficiency $\eta$. We found that the conditional state
approaches the squeezed Fock state $S(r)\ket{1}$ when $\tau,\eta \to 1$,
i.e., in the limit of single-photon subtraction. For non-unit IPS
transmissivity and efficiency, the conditioned state remains close to the
target state, {\em i.e.} shows a high fidelity for a wide range of experimental
parameters. The nonclassicality of the conditional state is also
investigated and a nonclassicality threshold on the IPS parameters 
is derived.
\end{abstract}
\section{Introduction}\label{s:1}
Beam splitters (BS) and avalanche photodetectors (APDs) play a fundamental
role in quantum information processing. These key elements, among the other
applications, can be used in order to generate non-Gaussian states
from Gaussian ones \cite{opatr:PRA:61,coch:PRA:65,ips:PRA:67,ips:PRA:70}
and to distill continuous-variable entanglement \cite{eis:AnnPhys:04}.
\par
In this paper we focus our attention on the output state recently obtained
experimentally by J.~Wenger {\em et al.}~\cite{wenger:PRL:04} by means of
photon subtraction on a squeezed vacuum state $S(r)\ket{0}$, $S(r)$ being
the squeezing operator. More precisely, when a Gaussian state, such as
$S(r)\ket{0}$, is mixed with the vacuum at a beam splitter and, then,
on/off photodetection is performed on the reflected beam, an unknown number
of photons is subtracted from the input state and the output state is no
longer Gaussian, i.e., the input state is de-Gaussified: this is due to the
fact that the positive operator valued measure (POVM) describing the APD is
non-Gaussian.  Since the actual number of detected photons cannot be
resolved by the APD, in \cite{ips:PRA:67} we referred to this process as to
inconclusive photon subtraction (IPS). In general the IPS process can be
characterized by two parameters: the beam splitter transmissivity $\tau$
and the quantum efficiency $\eta$ of the APD. As we will see, the
conditional output state obtained by IPS on a squeezed vacuum is close to
the squeezed Fock state $S(r)\ket{1}$, which is otherwise 
difficult to produce by
Hamiltonian processes. For this reason, we address IPS as an effective
resource to generate those squeezed Fock states.  
We find that the IPS conditional state reduces to
$S(r)\ket{1}$ in the limit $\tau,\eta\to 1$, whereas for different values
of the transmissivity and of the quantum efficiency it remains close to
this target state, 
showing a high fidelity for a wide range of the parameters. Finally, since
the IPS state obtained from the squeezed vacuum is, in general, non
classical and mixed, we study how the purity and the nonclassical depth of
the IPS state depend on $\tau$, $\eta$, and on the input squeezing
parameter $r$.
\par
The paper is structured as follows: in Section \ref{s:IPS} we review the
main elements of the IPS process on a single mode of radiation. The 
fidelity between the IPS conditional state and the squeezed Fock state
$S(z)\ket{1}$, as well as its purity are then investigated in Section \ref{s:fid},
whereas section \ref{s:nonclass} is devoted to the analysis of the
nonclassicality of the IPS state. Finally, Section \ref{s:remarks}
closes the paper with some concluding remarks.

\section{The inconclusive photon subtraction process}\label{s:IPS}
\begin{figure}
\begin{center}
\includegraphics[scale=.9]{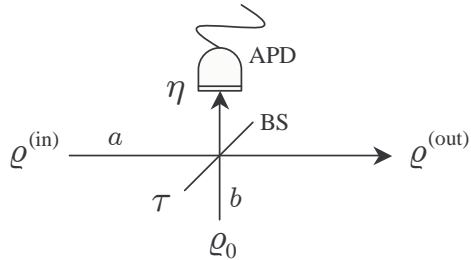}
\end{center}
\vspace{-.7cm}
\caption{\label{f:IPS:scheme} Scheme of the IPS process: the input state
$\varrho^{(\rm in)}$ is mixed with the vacuum state $\varrho_{0} =
\ket{0}\bra{0}$ at a beam splitter (BS) with transmissivity $\tau$; then,
avalanche photodetection (APD) with quantum efficiency $\eta$ is performed
on the reflected beam. When the detector clicks we obtain the IPS state
$\varrho^{(\rm out)}$.}\label{f:scheme}
\end{figure}
The scheme of the inconclusive photon subtraction (IPS) process is
sketched in figure \ref{f:scheme}.  An input state $\varrho^{(\rm in)}$ is
mixed with the vacuum state $\varrho_{0} = \ket{0}\bra{0}$ at a
beam splitter (BS) with transmissivity $\tau$ and, then, on/off avalanche
photodetection (APD) with quantum efficiency $\eta$ is performed on the
reflected beam. Since the APD con only distinguish the presence from the
absence of light, this measurement is {\em inconclusive}, namely does not
resolve the number of the detected photons. In this way, when the detector
clicks, an unknown number of photon is {\em subtracted} from the initial
state and we obtain the IPS state $\varrho^{(\rm out)}$. Since the whole
process is characterized by $\tau$ and $\eta$, we will refer to them also
as IPS transmissivity and IPS quantum efficiency.
\par
If the input state of the mode $a$ is the squeezed vacuum state
$\varrho^{(\rm in)}_r = \ket{0,r}\bra{0,r}$, where $\ket{0,r} =
S(r)\ket{0}$, $S(r) = \exp\{ \frac12 r ({a^{\dag}}^2 - a^2) \}$ being the
squeezing operator (for the sake of the simplicity, without lack of
generality, we can assume $r$ as real), its (Gaussian) characteristic
function $\chi^{(\rm in)}_r(\bmLambda_a) \equiv \chi[\varrho^{(\rm
in)}_{r}](\bmLambda_a)$ reads
\begin{equation}
\chi^{(\rm in)}_r(\bmLambda_a) =
\exp\left\{ -\mbox{$ \frac12 $} \bmLambda_a^T \bmsigma_r\, \bmLambda_a \right\}
\end{equation}
where $\bmLambda = (\sfx_a, \sfy_a)^T$, $(\cdots)^{T}$ being the
transposition operation, and
\begin{equation}
\bmsigma_r = \frac12 \left(
\begin{array}{cc}
\cosh r + \sinh r & 0\\
0 & \cosh r - \sinh r
\end{array}
\right)\,,
\end{equation}
is the covariance matrix. Analogously, the vacuum state $\varrho_0 =
\ket{0}\bra{0}$ of the mode $b$ is described by the (Gaussian)
characteristic
function
\begin{equation}
\chi_{0}(\bmLambda_b) \equiv
\chi[\varrho_0](\bmLambda_b) =
\exp\left\{ -\mbox{$ \frac12 $} \bmLambda_b^T \bmsigma_0\, \bmLambda_b
\right\}\,,
\end{equation}
where $\bmsigma_0 = \frac12 \mathbbm{1}_2$,
$\mathbbm{1}_2$ being the $2 \times 2$ identity matrix. Since the
initial two-mode state $\varrho^{(\rm in)}_r \otimes \varrho_0$ is Gaussian,
under the action of the BS its $4 \times 4$ covariance matrix 
\begin{equation}
\bmsigma_{\rm in} = \left(
\begin{array}{c|c}
\bmsigma_r & {\boldsymbol 0} \\
\hline
{\boldsymbol 0} & \bmsigma_0
\end{array}
\right)\,
\end{equation}
transforms as follows \cite{FOP:napoli:05}
\begin{equation}
\bmsigma_{\rm in} \rightsquigarrow \bmsigma' \equiv
{\bmS}_{\rm BS}^T\, \bmsigma_{\rm in}\,{\bmS}_{\rm BS} \equiv
\left(
\begin{array}{c|c}
\bmA & \bmC\\
\hline
\bmC^T & \bmB
\end{array}
\right)\,,
\end{equation}
where $\bmA$, $\bmB$, and $\bmC$ are $2 \times 2$ matrices and
\begin{equation}
\bmS_{\rm BS} = 
\left(
\begin{array}{c|c}
\sqrt{\tau}\, \mathbbm{1}_2 & \sqrt{1-\tau}\, \mathbbm{1}_2 \\
\hline
-\sqrt{1-\tau}\, \mathbbm{1}_2  & \sqrt{\tau}\, \mathbbm{1}_2
\end{array}
\right)\,,
\end{equation}
is the symplectic transformation associated to the evolution operator
of the BS. Now, the on/off photodetector with quantum
efficiency $\eta$ can be described by the POVM
$\left\{ \Pi_{\rm off}(\eta),\Pi_{\rm on}(\eta)\right\} $, with 
\begin{equation}
\Pi_{\rm off}(\eta) = \sum_{k=0}^{\infty}(1-\eta)^k\ket{k}\bra{k},\quad
\Pi_{\rm on}(\eta) = \mathbbm{I} - \Pi_{\rm off}(\eta)\,,
\end{equation}
which corresponds to the characteristic functions
\begin{eqnarray}
\chi[\Pi_{\rm off}(\eta)](\bmLambda) \equiv
\chi^{(\rm off)}_{\eta}(\bmLambda) =  \frac{1}{\eta}
\exp\left\{ -\mbox{$ \frac12 $} \bmLambda^T \bmsigma_{\rm M}\, \bmLambda
\right\}\,,\\
\chi[\Pi_{\rm on}(\eta)](\bmLambda) \equiv
\chi^{(\rm on)}_{\eta}(\bmLambda) = 2 \pi \delta^{(2)}(\bmLambda) -
\chi^{(\rm off)}_{\eta}(\bmLambda)\,,
\end{eqnarray}
respectively, $\delta^{(2)}(\bmLambda)$ being the 2-dim Dirac's delta
function, and
\begin{equation}
\bmsigma_{\rm M} = \frac{2-\eta}{2\eta}\,\mathbbm{1}_2\,.
\end{equation}
The probability of a click in the detector is then given by
\cite{FOP:napoli:05}
\begin{eqnarray}
\fl p_{\rm on}(r,\tau,\eta) &=
{\rm Tr}_{ab}[\varrho'_{r,\tau}\, \mathbbm{I}\otimes\Pi_{\rm on}(\eta)]\\
\fl &=\frac{1}{(2\pi)^2} \int_{\mathbbm{R}^4}
d^2\bmLambda_a\,d^2\bmLambda_b\,
\chi[\varrho'_{r,\tau}](\bmLambda_a,\bmLambda_b)\,
\chi[\mathbbm{I}](-\bmLambda_a)\,\chi^{(\rm on)}_{\eta}(-\bmLambda_b)\\
\fl &= 1 - \left(\eta\sqrt{{\rm Det}[\bmB + \bmsigma_{\rm
M}]}\right)^{-1}
= 1 - \left(\sqrt{ 1+(1-\tau_{\rm eff}^2)\sinh^{2}r}\right)^{-1}\,,
\label{click:prob}
\end{eqnarray}
where $\chi[\varrho'_{r,\tau}](\bmLambda_a,\bmLambda_b)$ is the two-mode
characteristic function associated to the state
$\varrho'_{r,\tau} \equiv U_{\rm BS}\,\varrho^{(\rm in)}_r\otimes\varrho_0\,
U_{\rm BS}^{\dag}$, $\chi[\mathbbm{I}](\bmLambda) = 2\pi
\delta^{(2)}(\bmLambda)$, and $\tau_{\rm eff}\equiv \tau_{\rm
eff}(\tau,\eta) = 1 - \eta (1 -\tau)$.
Note that when $\tau_{\rm eff} \to 1$, the probability (\ref{click:prob})
can be approximated at the first order in $\tau_{\rm eff}$ as follows
\begin{equation}
p_{\rm on}(r,\tau,\eta) = (1-\tau_{\rm eff}) \sinh^2 r
+o\left[ (1-\tau_{\rm eff})^2 \right]\,.
\end{equation}
Finally, the output state
\begin{equation}
\varrho^{(\rm out)}_{r,\tau,\eta} =
\frac{{\rm Tr}_b [\varrho'_{r,\tau}\,
\mathbbm{I}\otimes\Pi_{\rm on}(\eta)]}{p_{\rm on}(r,\tau,\eta)}\,,
\end{equation}
conditioned to a click of the on/off photodetector, has the following
characteristic function $\chi^{(\rm out)}_{r,\tau,\eta}(\bmLambda_a) \equiv
\chi[\varrho^{(\rm out)}_{r,\tau,\eta}](\bmLambda_a)$: 
\begin{eqnarray}
\fl \chi^{(\rm out)}_{r,\tau,\eta}(\bmLambda_a) &=
\frac{1}{2\pi\,p_{\rm on}(r,\tau,\eta)} \int_{\mathbbm{R}^2}
d^2\bmLambda_b\, \chi[\varrho'_{r,\tau}](\bmLambda_a,\bmLambda_b)\,
\chi^{(\rm on)}_{\eta}(-\bmLambda_b)\\
\fl &= \frac{1}{p_{\rm on}(r,\tau,\eta)} \left\{
\exp\left\{-\mbox{$\frac12$}\bmLambda_a^T\, \bmSigma_1\,
\bmLambda_a \right\} -
\frac{ \exp\left\{-\mbox{$\frac12$}\bmLambda_a^T\, \bmSigma_2\,
\bmLambda_a \right\}}{\eta\sqrt{{\rm Det}[\bmB + \bmsigma_{\rm M}]}}
\right\}\,,\label{chi:out:cartesian}
\end{eqnarray}
with $\bmSigma_1 = \bmA$ and $\bmSigma_2 = \bmA -\bmC (\bmB +
\bmsigma_{\rm M})^{-1} \bmC^T$. Note that the output state is no longer a
Gaussian state, namely its characteristic function is no longer Gaussian:
for this reason the IPS process is also referred to as {\em
de-Gaussification} process \cite{wenger:PRL:04}.
\par
\begin{figure}[tb]
\vspace{-1.5cm}
\setlength{\unitlength}{.9mm}
\begin{center}
\begin{picture}(70,140)(0,0)
\put(4,0){\includegraphics[width=60\unitlength]{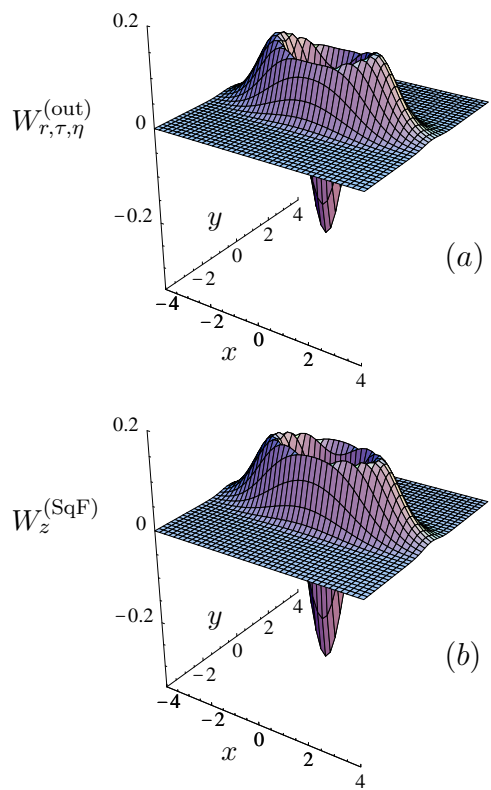}}
\put(-9,102){\small $W^{(\rm out)}_{r,\tau,\eta}$}
\put(22,68){\small $x$}
\put(20,88){\small $y$}
\put(55,82){$(a)$}
\put(-9,43){\small $W^{(\rm SqF)}_{z}$}
\put(22,9){\small $x$}
\put(20,29){\small $y$}
\put(55,23){$(b)$}
\end{picture}
\end{center}
\vspace{-.5cm}
\caption{(a) Plot of the Wigner function $W^{(\rm out)}_{r,\tau,\eta}(x,y)$
with $r=0.5$, $\tau=0.90$, and $\eta=0.80$; (b) plot of the Wigner function
$W^{(\rm SqF)}_{z}(x,y)$ of the state $S(z)\ket{1}$ with squeezing
parameter $z=0.5$.} \label{f:IPSvsFock}
\end{figure}
\begin{figure}[tb]
\vspace{-2cm}
\setlength{\unitlength}{1.1mm}
\begin{center}
\begin{picture}(70,140)(0,0)
\put(4,0){\includegraphics[width=60\unitlength]{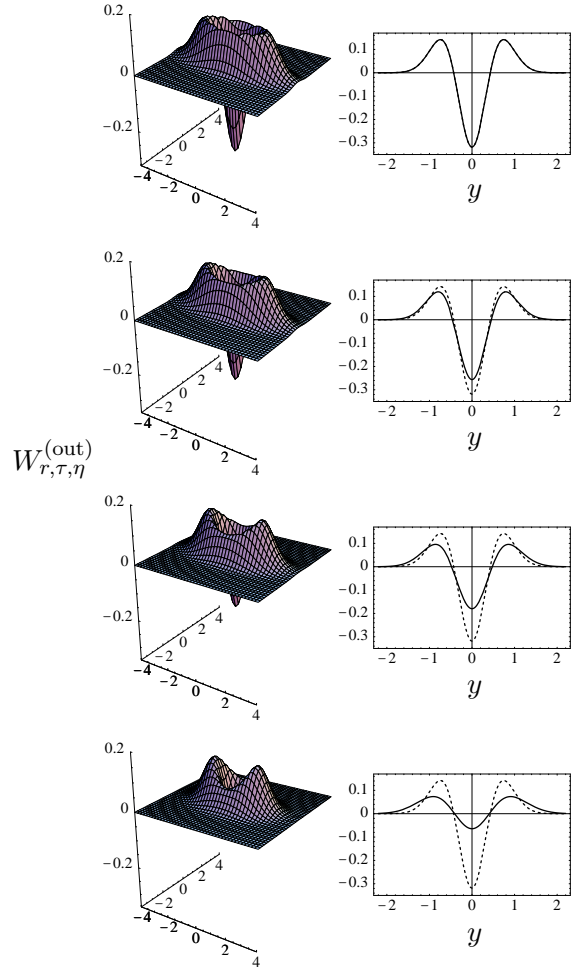}}
\put(-5,65){\small $W^{(\rm out)}_{r,\tau,\eta}$}
\put(50,98){\small $y$}
\put(50,68){\small $y$}
\put(50,38){\small $y$}
\put(50,8){\small $y$}
\end{picture}
\end{center}
\vspace{-.5cm}
\caption{Plots of the Wigner function $W^{(\rm out)}_{r,\tau,\eta}(x,y)$
with $r=0.5$, $\eta=0.80$ and different values of the BS transmissivity
$\tau$: from top to bottom $\tau = 0.99$, $0.9$, $0.75$, and $0.50$. The
solid lines of the plots on the right refer to $W^{(\rm
out)}_{r,\tau\eta}(0,y)$ whereas the dashed lines are $W^{(\rm
SqF)}_{z}(0,y)$ of the state $S(z)\ket{1}$ with squeezing parameter
$z=0.5$. Note that when $\tau = 0.99$ the two lines overlap.
$y$ is the squeezed coordinate.} \label{f:IPSprofile}
\end{figure}
\begin{figure}
\vspace{0.2cm}
\setlength{\unitlength}{1mm}
\begin{center}
\begin{picture}(70,80)(0,0)
\put(4,0){\includegraphics[width=60\unitlength]{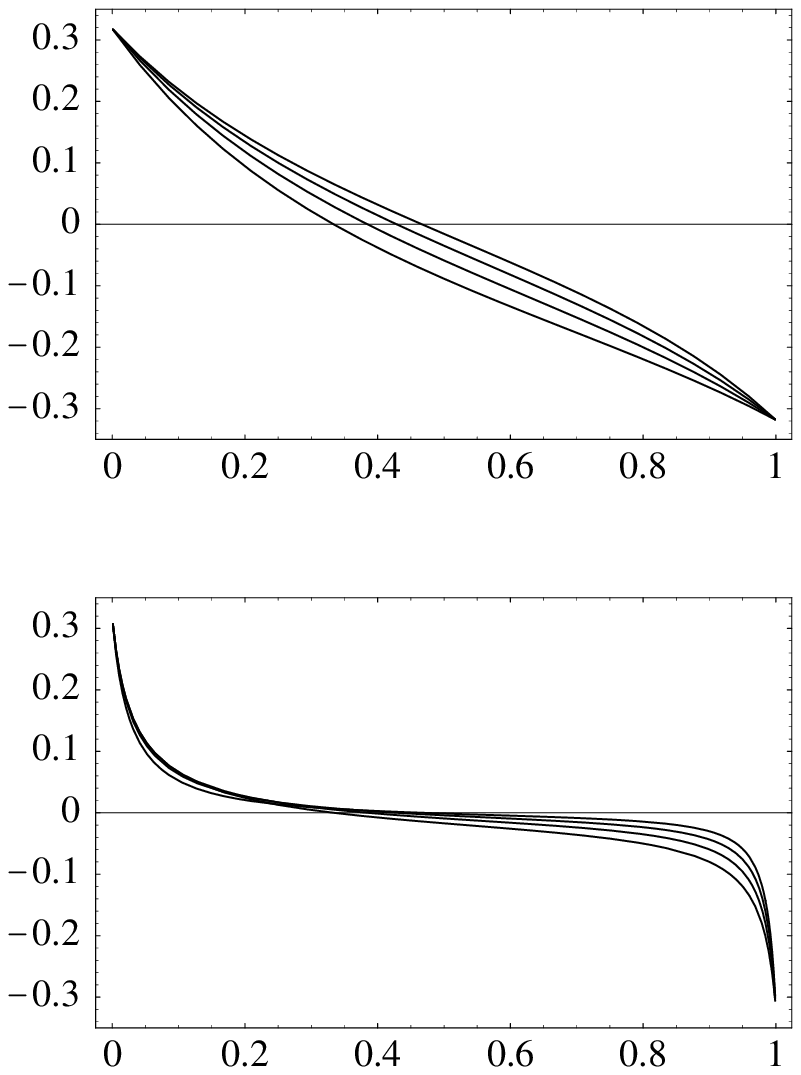}}
\put(55,73.5){(a)}
\put(55,31){(b)}
\put(35,-1.5){$\tau$}
\put(35,41){$\tau$}
\put(-13,63.5){\small $W^{(\rm out)}_{r,\tau,\eta}(0,0)$}
\put(-13,21){\small $W^{(\rm out)}_{r,\tau,\eta}(0,0)$}
\end{picture}
\end{center}
\vspace{-.2cm}
\caption{Plots of $W^{(\rm out)}_{r,\tau,\eta}(0,0)$ with (a) $r=0.5$ and
(b) $r=2.0$ as a function of $\tau$ and different values of $\eta$: from
bottom to top $\eta=1.0$, $0.75$, $0.50$, and $0.25$.
The value of the function is mainly affected
by $\tau$.} \label{f:IPSeta}
\end{figure}

In general, a Gaussian state described by the characteristic function [in
Cartesian notation, namely $\bmLambda = (\sfx , \sfy)^T$]
\begin{equation}
\chi(\bmLambda) =
\exp\left\{ -\mbox{$ \frac12 $} \bmLambda^T \bmsigma\, \bmLambda \right\}
\end{equation}
with covariance matrix
\begin{equation}
\bmsigma = \left(
\begin{array}{cc}
{\sf a} & {\sf c}\\
{\sf c} & {\sf b}
\end{array}
\right)\,,
\end{equation}
can be also written in the complex notation as follows:
\begin{equation}
\chi(\lambda) =
\exp\left\{ -\calA |\lambda|^2 -\calB \lambda^2 -\calB^* {\lambda^*}^2
\right\}\,,
\end{equation}
with
\begin{equation}\label{car2com}
\calA = \mbox{$\frac12$}({\sf a} + {\sf b})\,, \quad
\calB = \mbox{$\frac14$}({\sf b} - {\sf a} + 2 i {\sf c})\,,
\end{equation}
where we introduced the complex number
$\lambda = \frac{1}{\sqrt{2}}({\sf x} + i {\sf y})$.
In this way, the characteristic function
(\ref{chi:out:cartesian}) can be written as follows:
\begin{equation}\label{chi:out:complex}
\fl \chi^{(\rm out)}_{r,\tau,\eta}(\lambda) =
\frac{ \exp\left\{-\calA_1 |\lambda|^2 -\calB_1 \lambda^2
-\calB_1^* {\lambda^*}^2 \right\}}{p_{\rm on}(r,\tau,\eta)}
-\frac{ \exp\left\{ -\calA_2 |\lambda|^2 -\calB_2 \lambda^2
-\calB_2^* {\lambda^*}^2
\right\}}{p_{\rm on}(r,\tau,\eta)\,
\eta \sqrt{{\rm Det}[\bmB + \bmsigma_{\rm M}]}}\,,
\end{equation}
where $\calA_k$ and $\calB_k$ are refers to the covariance matrix
$\bmSigma_k$, $k=1,2$ respectively. 
Finally, using the definition
\begin{equation}
W[\varrho](\alpha) = \frac{1}{\pi^2} \int_{\mathbbm{C}} d^2 \lambda\,
\chi[\varrho](\lambda)\, \exp\left\{\lambda^* \alpha -
\alpha^* \lambda \right\}\,,
\end{equation}
which relates the Wigner function $W[\varrho](\alpha)$ of a state $\varrho$
to its characteristic function $\chi[\varrho](\lambda)$, one can obtain the
Wigner function $W^{(\rm out)}_{r,\tau,\eta}(\alpha) \equiv
W[\varrho^{(\rm out)}_{r,\tau,\eta}](\alpha)$. 
As for the characteristic function, to
pass from the complex, $W[\varrho](\alpha)$, to the Cartesian notation,
$W[\varrho](x,y)$, one should put $\alpha = \frac{1}{\sqrt{2}}(x+iy)$
\cite{FOP:napoli:05}. In figure \ref{f:IPSvsFock} (a) we report $W^{(\rm
out)}_{r,\tau,\eta}(x,y)$ for fixed $r$, $\tau$, and $\eta$: as it is
apparent from the plot the Wigner function is not Gaussian, and may
assume negative values
\cite{wenger:PRL:04}. In Section \ref{s:nonclass} we will investigate this
effect by analyzing the nonclassicality of the conditioned state. In figure
\ref{f:IPSvsFock} (b) we show the Wigner function $\chi^{(\rm
SqF)}_{z}(x,y)$ associated to the squeezed Fock state $\varrho^{(\rm
SqF)}_{z}= S(z)\ket{1} \bra{1}S^{\dag}(z)$, whose characteristic function
$\chi^{(\rm SqF)}_{z}(\lambda)\equiv\chi[\varrho^{(\rm SqF)}_{z}](\lambda)$
reads (we assume $z$ as real)
\begin{equation}\label{chi:squeezed:complex}
\fl \chi_{z}^{(\rm SqF)}(\lambda) =
\left[ 1 - 2 \left( \calA_0 |\lambda|^2 + \calB_0 \lambda^2 +
\calB_0^* {\lambda^*}^2 \right) \right]\,
\exp\left\{ -\calA_0 |\lambda|^2 - \calB_0 \lambda^2 -
\calB_0^* {\lambda^*}^2  \right\}\,,
\end{equation}
with $\calA_0 = 2 (\cosh^2 z + \sinh^2 z)$ and
$\calB_0 = -2 \cosh z\,  \sinh z$. 
Since the Wigner functions of the IPS squeezed vacuum and of the squeezed
number state are quite similar, one can think of using the IPS process to
produce the state $\varrho^{(\rm SqF)}_{r}$; motivated by this
consideration, in the next section we will analyses the fidelity between
this states.
Figure \ref{f:IPSprofile} shows $W^{(\rm out)}_{r,\tau,\eta}(x,y)$
with fixed $r$ and $\eta$ and different values of the IPS transmissivity
$\tau$; the plots on the right of the same figure compare the
$W^{(\rm out)}_{r,\tau\eta}(0,y)$ (solid lines) with
$W^{(\rm SqF)}_{r}(0,y)$ (dashed line).
Finally, the effect of the quantum efficiency $\eta$ on the output state is
shown in figure \ref{f:IPSeta}, where we plot as reference the value of the
Wigner function $W^{(\rm out)}_{r,\tau,\eta}$ at the center of the complex
plane as a function of the transmissivity $\tau$ and different values of
$\eta$: we can see that the main effect on the output state is due to $\tau$.

\section{Fidelity and purity}\label{s:fid}
The fidelity between the pure state $\varrho^{(\rm SqF)}_{z}$ and the
IPS state $\varrho^{(\rm out)}_{r,\tau,\eta}$ is defined as follows:
\begin{eqnarray}\label{fid:SqFIPS}
F_{\tau,\eta}(z,r) &= {\rm Tr}[\varrho^{(\rm SqF)}_{z}\,
\varrho^{(\rm out)}_{r,\tau,\eta}]\\
&=\frac{1}{2\pi}\int_{\mathbbm{R}^2} d^2 \bmLambda\,
\chi^{(\rm SqF)}_{z}(\bmLambda)\,
\chi^{(\rm out)}_{r,\tau,\eta}(-\bmLambda)\,,\\
&=\frac{1}{p_{\rm on}(r,\tau,\eta)}\left\{
{\cal F}_1 -\frac{{\cal F}_2}{\eta\sqrt{{\rm Det}[\bmB +
\bmsigma_{\rm M}]}}\,,
\right\}
\end{eqnarray}
where
\begin{equation}
{\cal F}_k =
\frac{\displaystyle
\calA_k^2 - \calA_0^2 - 4 (\calB_k^2 - \calB_0^2)
}{[(\calA_0 + \calA_k)^2-4(\calB_0 + \calB_k)^2]^{3/2}}
\end{equation}
and $\calA_h$ and $\calB_h$, $h = 0,1,2$, have been introduced in equations
(\ref{chi:squeezed:complex}) and (\ref{chi:out:complex}), respectively. The
analytic expression of $F_{\tau,\eta}(z,r)$ is quite cumbersome, but, on the
other hand, we can draw some interesting consideration by addressing its
expansion at the first order in the transmissivity $\tau$ when $\tau \to 1$
and $\eta = 1$, namely
\begin{eqnarray}\label{fid:expansion}
\lo{ F_{\tau,1}(z,r) = } 
\frac{1}{\cosh^3(r - z)}\nonumber\\
- \Bigg[ \frac{ 9 \cosh(r + z)
- 3 \cosh(3 r - z)}{8\cosh(r-z)} - \frac14 \Bigg] (1-\tau)
+ o\left[ (1-\tau)^2 \right]\,.
\end{eqnarray}
In fact, from the expansion (\ref{fid:expansion}) we conclude that the
maximum of the fidelity is achieved when $z = r$.
\par
\begin{figure}
\vspace{-1cm}
\setlength{\unitlength}{1mm}
\begin{center}
\begin{picture}(70,50)(0,0)
\put(4,0){\includegraphics[width=60\unitlength]{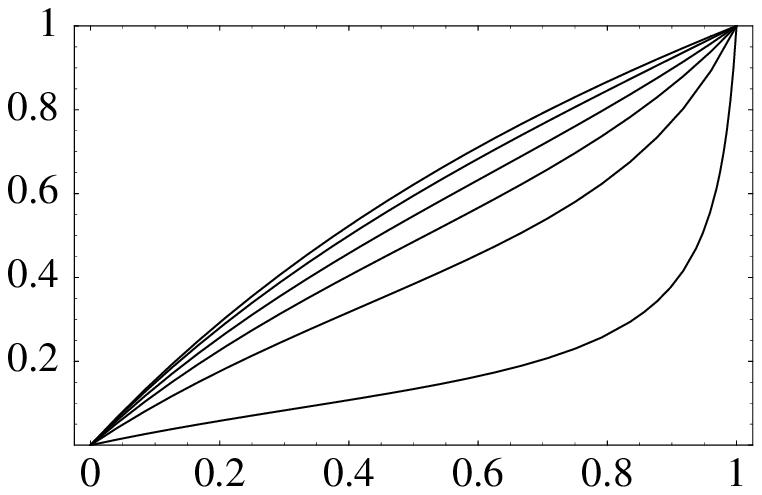}}
\put(40,-3){$\tau$}
\put(-3,24.5){$F_{\tau,\eta}$}
\end{picture}
\end{center}
\caption{Plot of the fidelity $F_{\tau,\eta}(r)$ with $\eta = 0.80$ as a
function of the IPS transmissivity $\tau$ for different values of
$r$: from top to bottom $r = 0.1$, $0.3$, $0.5$, $0.7$, $1.0$, and $2.0$.}
\label{f:fidSqFvsIPS}
\end{figure}
\begin{figure}
\vspace{0.2cm}
\setlength{\unitlength}{1mm}
\begin{center}
\begin{picture}(70,50)(0,0)
\put(4,0){\includegraphics[width=60\unitlength]{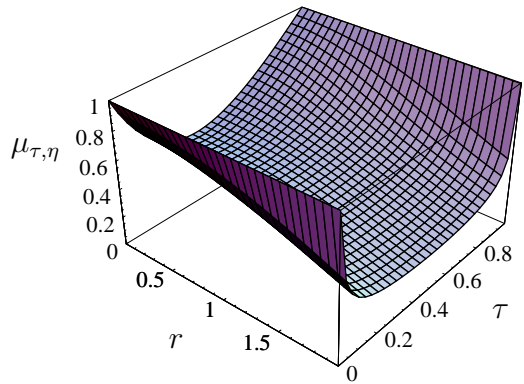}}
\put(17,6){$r$}
\put(60,10){$\tau$}
\put(-4,32){$\mu_{\tau,\eta}$}
\end{picture}
\end{center}
\vspace{-.5cm}
\caption{Plot of the purity $\mu_{\tau,\eta}(r)$ of the state
$\varrho^{(\rm out)}_{r,\tau,\eta}$. We set $\eta = 0.80$.} \label{f:purity}
\end{figure}

In figure \ref{f:fidSqFvsIPS} we plot $F_{\tau,\eta}(r) \equiv
F_{\tau,\eta}(r,r)$ as a function of the IPS transmissivity and for
different values of $r$. We can see that $F_{\tau,\eta}$ reaches it maximum
when the IPS transmissivity approaches $1$, namely in the single-photon
subtraction limit \cite{ips:PRA:67}. Moreover, when the squeezing parameter
$r$ increases the fidelity decreases: this is due to the increasing
(unknown) number of subtracted photons which reduces the {\em purity} of
the IPS state itself. In figure \ref{f:purity} we plot the purity
$\mu_{\tau,\eta}(r)$ of the IPS squeezed vacuum $\varrho^{(\rm
out)}_{r,\tau,\eta}$, defined as follows \cite{paris:purity}:
\begin{eqnarray}
\fl \mu_{\tau,\eta}(r) =
{\rm Tr}\left[(\varrho^{(\rm out)}_{r,\tau,\eta})^2\right]
= \pi \int_{\mathbbm{C}} d^2\alpha
\left[W^{(\rm out)}_{r,\tau,\eta}(\alpha)\right]^2\\
\lo= \frac{1}{2 p_{\rm on}(r,\tau,\eta)}\left\{
\frac{1}{\sqrt{\calA_1^2 - 4\calB_1^2}}+
\frac{1}{\eta^2\,{\rm Det}[\bmB+\bmsigma_{\rm M}]\,\sqrt{\calA_1^2 -
4\calB_1^2}}\right.\nonumber\\
\left.
-\frac{4}{\eta \sqrt{{\rm Det}[\bmB+\bmsigma_{\rm M}]}\,
\sqrt{(\calA_1 + \calA_2)^2 - 4(\calB_1+\calB_2)^2}}
 \right\}\,.
\end{eqnarray}

\section{Nonclassicality of the IPS squeezed vacuum state}\label{s:nonclass}
\begin{figure}
\vspace{0.2cm}
\setlength{\unitlength}{1mm}
\begin{center}
\begin{picture}(70,50)(0,0)
\put(4,0){\includegraphics[width=60\unitlength]{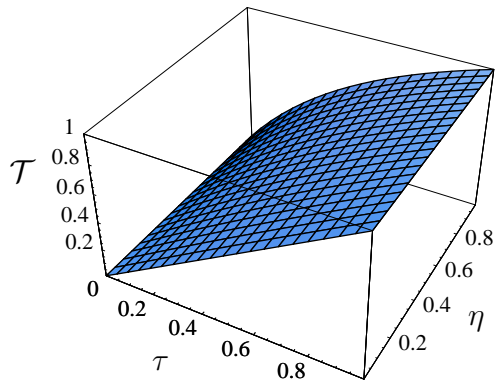}}
\put(18,4){$\tau$}
\put(60,10){$\eta$}
\put(-1,29){${\cal T}$}
\end{picture}
\end{center}
\vspace{-.5cm}
\caption{Plot of the nonclassical depth ${\cal T}(\tau,\eta)$ of the IPS
squeezed vacuum state.}
\label{f:NonDepth}
\end{figure}

As a measure of nonclassicality of the IPS state $\varrho^{(\rm
out)}_{r,\tau,\eta}$ we consider the {\em nonclassical depth}
\cite{lee:PRA:91}
\begin{equation}
{\cal T} = \frac{1-\overline{s}}{2}\,,
\end{equation}
$\overline{s}$ being the maximum $s$ for which the generalized
quasi-probability function
\begin{equation}\label{gen:wigner}
W_s(\alpha) = \frac{1}{\pi}\int_{\mathbb{C}}d^2\lambda\,
\chi(\lambda)\,
\exp\left\{ \mbox{$\frac12$}s + \lambda^* \alpha - \alpha^* \lambda
\right\}
\end{equation}
is a probability distribution, i.e.~positive semidefinite and non singular.
As a matter of fact, one has ${\cal T} = 1$ for number states and ${\cal T}
= 0$ for coherent states. Moreover, the nonclassical depth can be
interpreted as the minimum number of thermal photons which has to be added
to a quantum state in order to erase all the quantum features of the state
\cite{FOP:napoli:05, lee:PRA:91}. In the case of $\varrho^{(\rm
out)}_{r,\tau,\eta}$, we have [for the sake of simplicity we do not write
explicitly the dependence on $r$,$\tau$ and $\eta$ in the symbol
$W_{s}^{\rm (out)}(\alpha)$]
\begin{equation}\label{gen:wig:s}
W_{s}^{\rm (out)}(\alpha) = \frac{1}{p_{\rm on}(r,\tau,\eta)}\left\{ 
{\cal G}_{1}(\alpha) -
\frac{ {\cal G}_{2}(\alpha) }{\eta \sqrt{{\rm Det}[\bmB + \bmsigma_{\rm M}]}}
\right\}
\end{equation}
where we defined
\begin{equation}
{\cal G}_{k}(\alpha) = \frac{\displaystyle 2\, \exp\left\{
-\frac{2(2\calA_k -s)|\alpha|^2 + 4 \calB_k^* \alpha^2 + 4 \calB_k
{\alpha^*}^2}{(2\calA_k-s)^2-16 |\calB_k|^2}
\right\}}{\pi \sqrt{(2\calA_k-s)^2-16 |\calB_k|^2}}\, \,.
\end{equation}
At first we note that in order to have $W_{s}^{\rm (out)}(\alpha)$ 
normalizable, equation the following condition should be satisfied
\begin{equation}\label{cond:norm}
s \le 2 \calA_k\qquad (k=1,2)\,.
\end{equation}
Furthermore, since $W_{s}^{\rm (out)}(\alpha)$ is a difference between two
Gaussian functions with the center in the origin of the complex plane, one
can easily see that, in general, this function has a minimum in $\alpha =
0$ and that this minimum can be negative. For this reason and thanks to
other simple considerations about the symmetries of $W_{s}^{\rm
(out)}(\alpha)$ with respect to the point $\alpha = 0$, we can focus our
attention in the origin of the complex plane, obtaining this further
condition for the positivity:
\begin{equation}
{\cal G}_{1}(0) -
\frac{ {\cal G}_{2}(0) }{\eta\sqrt{{\rm Det}[\bmB + \bmsigma_{\rm M}]}}
\ge 0\,,
\end{equation}
which, together with the conditions (\ref{cond:norm}), brings to
\begin{equation}\label{maxs}
\overline{s}(\tau,\eta) = \frac{2-\eta-(4-\eta)\tau}{2-(1-\tau)\eta}\,,
\end{equation}
and, then, to the following expression for the nonclassical depth:
\begin{equation}\label{nondepth}
{\cal T}(\tau,\eta) = \frac{2\tau}{2-(1-\tau)\eta}\,.
\end{equation}
Since ${\cal T}(\tau,\eta) \ge 0$, the conditional state is nonclassical
for any non-zero value of the IPS transmissivity and efficiency.
Note that  equation (\ref{nondepth}) depends only on $\tau$ and $\eta$,
whereas it is independent on the squeezing parameter $r$.
Notice, however, the nonclassical depth does not measure the extension of
the negativity region, but only the presence of negative values. Therefore
it is not surprising that equation (\ref{nondepth}) does not depend on $r$.
We plot ${\cal T}(\tau,\eta)$ in figure \ref{f:NonDepth}. Since the usual
Wigner function is obtained when $s = 0$ in (\ref{gen:wigner}), from
equation (\ref{maxs}) we can see that $W^{(\rm out)}_{r,\tau,\eta}(\alpha)$
becomes semi-positive definite when $\tau = (2-\eta)/(4-\eta)$.

\section{Concluding remarks}\label{s:remarks}
We have analyzed in details the state obtained subtracting photons from the
squeezed vacuum by means of linear optics, namely using beam splitters and
avalanche photodetectors. We referred to the whole photon-subtraction
process as to inconclusive photon subtraction (IPS), since avalanche
photodetectors are not able to resolve the number of detected photons. We
found that the IPS conditional state obtained from a 
squeezed vacuum state is close to the squeezed Fock
state $S(r)\ket{1}$ and approaches this target state when only one photon is
subtracted, namely, using a high transmissivity beam splitter for the IPS.
Moreover, when the transmissivity and the quantum efficiency
are not unitary, the output state remains close to the target
state, showing a high fidelity for a wide range of the parameters. The
purity and the nonclassicality of the IPS squeezed vacuum state have been
also considered: we found that the relevant parameter is the
transmissivity $\tau$, while the IPS efficiency $\eta$ only slightly
affects the output state. We conclude that IPS, which was recently
experimentally implemented \cite{wenger:PRL:04}, can be effectively 
used to produce a nonclassical state such as the squeezed Fock state
$S(r)\ket{1}$, whose generation would be, otherwise, quite challenging.


\end{document}